\documentclass[aps,twocolumn]{revtex4}
\usepackage[dvips]{graphics,graphicx}
\usepackage{amsmath}
\begin{document}

\title{Effects of internal and external decoherence on the resonant transport and Anderson localization of fermionic particles in the tight-binding chain}
\author{Andrey~R.~Kolovsky}
\affiliation{Kirensky Institute of Physics, Federal Research Center KSC SB RAS, 660036, Krasnoyarsk, Russia}
\affiliation{School of Engineering Physics and Radio Electronics, Siberian Federal University, 660041, Krasnoyarsk, Russia}

\date{\today}
\begin{abstract}
We study effects of relaxation/decoherence processes on quantum transport of non-interacting Fermi particles across the tight-binding chain, where we distinguish between relaxation processes in the contacts (external decoherence) and those in the chain (internal decoherence). It is argued that  relaxation processes in the contacts can essentially modify the resonant transmission as compared to the Landauer theory. We also address quantum transport in disordered chains. It is shown that external decoherence reduces conductance fluctuations but does not alter the Anderson localization length. This is in strong contrast with the effect of internal decoherence which is found  to suppress the Anderson localization. 
\end{abstract}
\maketitle

\section{Introduction}
One finds phenomena of resonant transmission in many different physical systems, including the mesoscopic semiconductor devices where the electric current is greatly enhanced for certain values of the gate voltage. The origin of this effect is the enhanced transmission probability for the Bloch wave when its energy coincides with one of the device eigenenergies. Thus, if we fix the Bloch wave energy to the Fermi energy of electrons in the contacts/leads and change the gate voltage, the transmission probability will show a number of resonant peaks. However, in laboratory experiments one measures not the transmission probability  for the Bloch wave but the electric current, where the resonant peaks are broadened. As is known, there are several mechanisms for peak broadening: (i) a finite deference in the chemical potential of the contacts, (ii) a finite temperature, and (iii) quantum decoherence due to relaxation processes which usually present in the system. While the first two mechanisms are well studied, the last one is still under the active research \cite{Jin20,Visu22,Uchi22,Visu23}. 

Theoretically, there are two main approaches for describing the quantum transport in a mesoscopic device in the presence of relaxation processes. These are the Keldysh formalism \cite{Sieb16}, which can be viewed as further development of the Landauer-B\"uttiker \cite{Land57,Buet85,Datt95} and non-equilibrium Green's functions \cite{Datt95,Meir08,Ramm08,Haug08,Kame11} theories,  and the quantum master equation for the system density matrix \cite{Pros14,120,Land22}. For the paradigm model of quantum transport -- the tight-binding chain coupled at both ends to fermionic reservoirs --  the equivalence between the Landauer and master equation approaches was demonstrated in our recent work Ref.~\cite{prl}. We mention that from the side of the master equation approach this equivalence requires abandonment of the Markov approximation which leads to the integro-differential evolution equation \cite{122,129}. 

In the present contribution we analyze the effect of relaxation processes on the resonant transmission by using the master equation approach of Ref.~\cite{prl}  As compared to the relevant Keldysh's formalism based studies \cite{Jin20,Visu22,Uchi22,Visu23}, the new twist is that we also take into account relaxation processes in the contacts. These processes force the isolated contacts to relax towards the thermal equilibrium and are characterized by the rate $\gamma$ which is often referred to as the self-thermalization rate \cite{Nand15,Ales16,Borg16,105,115}. We shall show that this seemingly minor modification of the transport problem can essentially modify the resonant transmission as compared to the Landauer theory.

The diametrically opposed phenomenon to the resonant transmission is the Anderson localization in disordered chains. An nice feature of the master equation approach is its conceptional  and numerical simplicity that allows us to consider long chains needed to address the Anderson  localization. We analyze effects of both internal and external decoherence caused by relaxation processes in the chain and contacts, respectively. It is shown that external decoherence has a little effect on the Anderson localization. On the contrary, even a weak internal decoherence strongly affects the Anderson localization length.

\section{The model and numerical method}
\label{sec1}
To make the paper self-consistent we briefly summarize the approach and main results of our previous works \cite{prl,122,129}.

We consider the tight-binding chain of the length $L$ connected at both ends to the contacts via the reduced hopping matrix element $\epsilon\ll J$. The contacts are modeled by the tight-binding rings of the size $M$, which eventually tends to infinity, and we incorporate in the model the Lindblad operators which enforce the isolated rings ($\epsilon=0$) to relax to the thermal equilibrium with the relaxation rate $\gamma$.  Thus, on the formal level
we have 
\begin{equation}
\label{Ham_tot}
\widehat{{\cal H}}=\widehat{{\cal H}}_{\rm s}+\sum_{j=1,L}\left(
\widehat{{\cal H}}_{{\rm r},j}+\widehat{{\cal H}}_{{\rm c},j}\right) .
\end{equation}
Here, the subindex '${\rm s}$' stands for the chain (the system), the subindex '${\rm r}$' for the rings (the reservoirs), the subindex '${\rm c}$' for the coupling between the system and reservoirs, and the index $j=1$ or $j=L$ corresponds to the left and right reservoirs, respectively. The explicit form of the Hamiltonians in Eq.~(\ref{Ham_tot}) is the following
\begin{eqnarray}
\label{Ham_sys}
\widehat{{\cal H}}_{\rm s}=\sum_{\ell=1}^{L}\delta_\ell \hat{c}_{\ell}^{\dagger}\hat{c}_{\ell}
-\frac{J}{2}\sum_{\ell=1}^{L-1}\hat{c}_{\ell+1}^{\dagger}\hat{c}_{\ell} +{\rm h.c.} \;,\\
\label{Ham_res}
\widehat{{\cal H}}_{\rm r}=
-J\sum_{k=1}^{M}\cos\left(\frac{2\pi k}{M}\right) \hat{b}_{k}^{\dagger}\hat{b}_{k} \;,\\
\label{Ham_cou}
\widehat{{\cal H}}_{{\rm c}, j}=
-\frac{\epsilon}{2\sqrt{M}}\sum_{k=1}^M\hat{c}_{j}^{\dagger}\hat{b}_{k} +{\rm h.c.} \;.
\end{eqnarray}
Notice that we use the quasimomentum basis for the rings and the Wannier basis for the chain.  

The governing master equation for the total density matrix ${\cal R}(t)$ reads 
\begin{equation}
\label{Master_full}
\frac{\partial \widehat{{\cal R}}}{\partial t}=-\frac{i}{\hbar}[\widehat{{\cal H}}, \widehat{{\cal R}}]+
\gamma\sum_{j=1,L}\left(\widehat{{\cal L}}^{(g)}_{j}+\widehat{{\cal L}}^{(d)}_{j}\right) ,
\end{equation}
where the Lindblad relaxation operators have the form
\begin{eqnarray}
\label{drain}
\widehat{{\cal L}}^{(d)}_{j}=\sum_{k=1}^M\frac{\bar{n}_{k,j}-1}{2}
\left(\hat{b}_{k}^{\dagger}\hat{b}_{k}\widehat{\cal R }-2\hat{b}_{k}\widehat{\cal R }\hat{b}_{k}^{\dagger}
+\widehat{\cal R }\hat{b}_{k}^{\dagger}\hat{b}_{k} \right) ,\\
\label{gain}
\widehat{{\cal L}}^{(g)}_{j}=-\sum_{k=1}^M\frac{\bar{n}_{k,j}}{2}
\left(\hat{b}_{k}\hat{b}_{k}^{\dagger}\widehat{\cal R }-2\hat{b}_{k}^{\dagger}\widehat{\cal R }\hat{b}_{k}
+\widehat{\cal R }\hat{b}_{k}\hat{b}_{k}^{\dagger} \right) ,\\
\label{Fermi}
\bar{n}_{k,j}= \frac{1}{e^{-\beta_{j}[J\cos(2\pi k/M)+\mu_{j}]}+1} \;.
\end{eqnarray}

Since we consider the case of non-interacting fermions, we can derive from the master equation (\ref{Master_full}) the equation for the single particle density matrix (SPDM) in the closed form. We have
\begin{equation} 
\label{SPDM}
\frac{d\rho}{dt}=-i[H,\rho]-G*\rho +\gamma\rho_0 \;.
\end{equation} 
In this equation $\rho$ is the SPDM of the size $(M+L+M)\times(M+L+M)$, $H$ the single-particle version of the many-body Hamiltonian (\ref{Ham_tot}), $\rho_0$ the diagonal matrix with elements determined by the Fermi distribution (\ref{Fermi}), $G$ the relaxation matrix with elements proportional to $\gamma$,  
\begin{equation} 
\label{gamma}
{G}_{n,m}=\left\{
\begin{array}{lll}
0&,&(n,m)\in L\times L \\
\gamma&,&(n,m)\in M\times M \\
\gamma/2&,&(n,m)\in L\times M, M\times L 
\end{array}
\right. \;,
\end{equation} 
and the sign '$*$' denotes element-by-element product of two matrices. We mention that  we use the ordering of the density matrix elements where the corner blocks of the size $M\times M$ are the contact SPDMs and the central block of the size $L\times L$ is the chain SPDM. It is easy to see from Eqs.~(\ref{SPDM})-(\ref{gamma}) that for $\epsilon=0$ the contacts relax to the thermal equilibrium and, simultaneously, any correlations between the chain and contacts (i.e., the system and reservoirs) decay to zero.  We are  interested in the non-equilibrium steady-state  for $\epsilon\ne0$, which obviously  obeys the following algebraic equation
\begin{equation} 
\label{algebraic}
i[H,\rho]+G*\rho =\gamma\rho_0 \;.
\end{equation} 

The displayed algebraic equation (\ref{algebraic}) is the starting point for both theoretical and numerical analysis of the system. In theory we derive from Eq.~(\ref{algebraic}) an equation for the chain SPDM $\rho_s$, that involves the assumption $\epsilon\ll J$ together with the limit $M\rightarrow\infty$ and the limit $\Delta\mu \rightarrow 0$.  In numerics, where we solve Eq.~(\ref{algebraic}) numerically,  we are not restricted by the condition $\epsilon\ll J$   but have to watch convergence of the result as $M$ is increased.  In practice this is done by comparing $\rho_s$ (the central block of the total SPDM) calculated for different values of $M$. The empirical rule is that $M$ should be larger than $1/\gamma$ to get the convergence.  Knowing  $\rho_s$ with the desired accuracy we then calculate the current across the chain as 
\begin{equation} 
\label{current}
j=\frac{1}{L-1}{\rm Tr}[\hat{j} \rho_s] \;,
\end{equation} 
(here $\hat{j}$ is the two-diagonal matrix of the single-particle current operator) and the systems conductance 
\begin{equation} 
\label{conductance}
\sigma(\delta)=\lim_{\Delta\mu\rightarrow0} \frac{j(\delta,\Delta\mu)}{\Delta\mu} \;.
\end{equation} 
We also find convenient to incorporate the latter limit in the limit $M\rightarrow\infty$. In the other words,  we proportionally decrease $\Delta\mu$
when increasing the ring size $M$.

In what follows, for the sake of simplicity, we shall assume zero temperatures of the contacts. Also,  from now on we use dimensionless parameters where the hopping matrix element $J$ is the energy unit. Thus, $\epsilon=0.1$ or $\gamma=0.1$ imply that the coupling constant and relaxation rate times the Planck constant are one tenth of the hopping energy.

\section{Resonant transmission and Anderson's localization}

The numerical analysis of the system conductance on the basis of the algebraic equation (\ref{algebraic}) reveals two asymptotic results which agrees very well with the analytical estimates. Namely, in the limit $\gamma\rightarrow0$ the conductance is given by the Landauer equation
\begin{equation} 
\label{Landauer}
\sigma(\delta)=\frac{1}{2\pi} |t(\delta)|^2
\end{equation} 
where $1/2\pi$ is the conductance quantum $e^2/h$ in our dimensionless units and $|t(\delta)|^2$ is the transmission probability which one finds by solving the scattering problem for the Bloch wave.  Eq.~(\ref{Landauer}) can be elaborated further  by assuming the week coupling limit $\epsilon\ll 1$. Then we have  
\begin{equation} 
\label{transmission}
\sigma(\delta) \approx \sum_{n=1}^L \frac{\Gamma_n^2}{\Gamma_n^2+(\delta-E_n)^2} \;, \quad  \Gamma_n =\frac{\alpha_n \epsilon^2}{2} \;,
\end{equation} 
where $E_n$ are the eigenvalues of the isolated chain and $\alpha_n$ are determined by the eigenfunctions $\psi_n$ of the isolated chain as 
\begin{equation} 
\label{alpha}
\alpha_n=| \langle \psi_n  | \langle \ell=1\rangle\langle \ell=L | \psi_n\rangle  |  \;.    
\end{equation} 
(Notice that $\sum_n \alpha_n=1$.)  It follows from the displayed equation that the width of conductance peaks is proportional to $\epsilon^2$ and all peals have the same height $1/2\pi$.
\begin{figure}[t]
\includegraphics[width=8.5cm,clip]{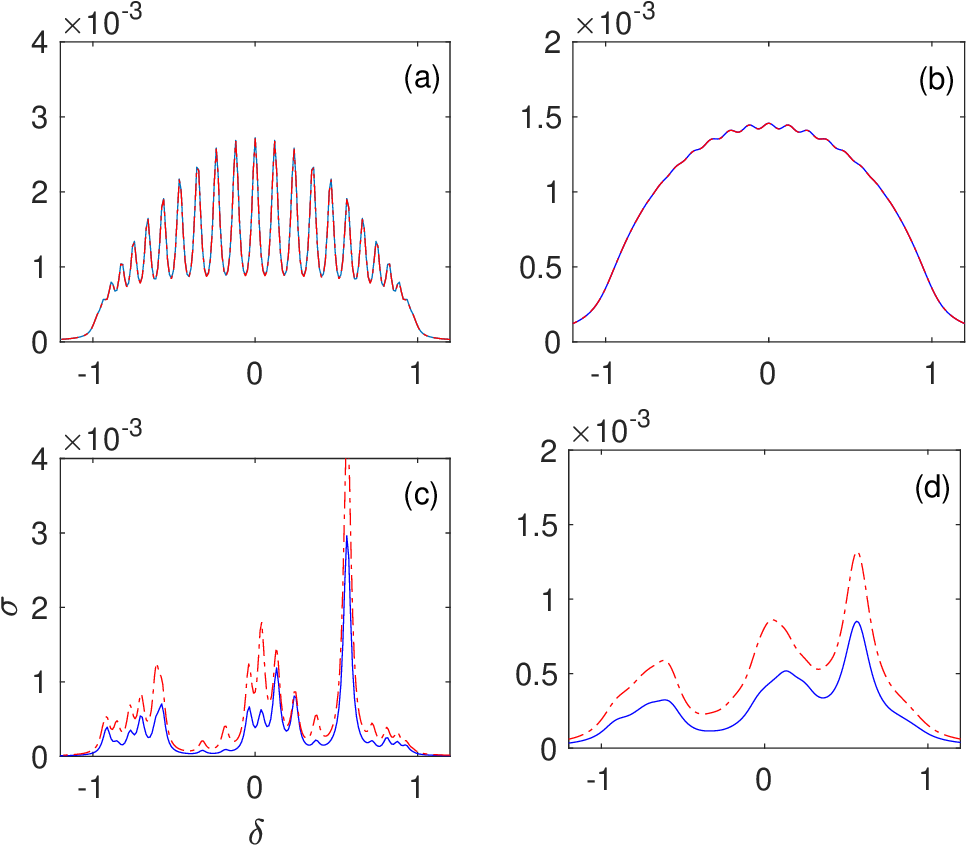}
\caption{Conductance of the chain $L=25$ for $\gamma=0.05$, left column, and $\gamma=0.2$, right column. The other parameters are  $\epsilon=0.1$, $\mu=0$, and  the disorder strength $\xi=0$, upper row, and $\xi=0.4$, lower row. The dash-dotted line is Eq.~(\ref{combined}).}
\label{fig5}
\end{figure}

In the opposite limit of large $\gamma$  the width of conductance peaks is determined by the parameter $\gamma$  while the peaks heights are proportional to $\epsilon^2$,  
\begin{equation} 
\label{Lorentz_sum}
\sigma(\delta) \approx \frac{\epsilon^2}{4\pi} \sum_{n=1}^L \alpha_n \frac{(\gamma/2)}{(\gamma/2)^2+(\delta-E_n)^2} \;.
\end{equation} 
Notice that the estimate (\ref{Lorentz_sum}) is obtained under the assumption $\epsilon^2\ll \gamma$ and, hence, the peak heights are always smaller than $1/2\pi$.   It is easy to show that Eq.~ (\ref{Lorentz_sum}) and Eq.~(\ref{Landauer}) can be combined into the single equation
\begin{equation} 
\label{combined}
\sigma(\delta) \approx \frac{1}{2\pi} \sum_{n=1}^L \frac{\Gamma_n(\Gamma_n+\gamma/2)}{(\Gamma_n+\gamma/2)^2+(\delta-E_n)^2} \;, 
\end{equation} 
which interpolates between the limits $\gamma\ll \epsilon^2$ and $\gamma\gg \epsilon^2$. From the side of the Landauer approach Eq.~ (\ref{combined}) can be viewed as broadening of the resonant peaks due to the partial decoherence of the carries transporting states in the chain  caused by the contact dissipative  dynamics \cite{prl}.

We compared the estimate (\ref{combined}) with the exact numerical results for different chain length $L$ and found an excellent agreement. As two examples, Fig.~\ref{fig5}(a,b) show the conductance  of the chain $L=25$ for $\gamma=0.05$, where the resonant peaks are partially resolved, and $\gamma=0.2$, where peaks merge together resulting in the universal dependence
\begin{equation}
\label{semicircle} 
\sigma(\delta)\approx \frac{\epsilon^2}{2\pi} \sqrt{J^2-\delta^2} \;,
\end{equation} 
which is independent of $L$. 
\begin{figure}[b]
\includegraphics[width=8.5cm,clip]{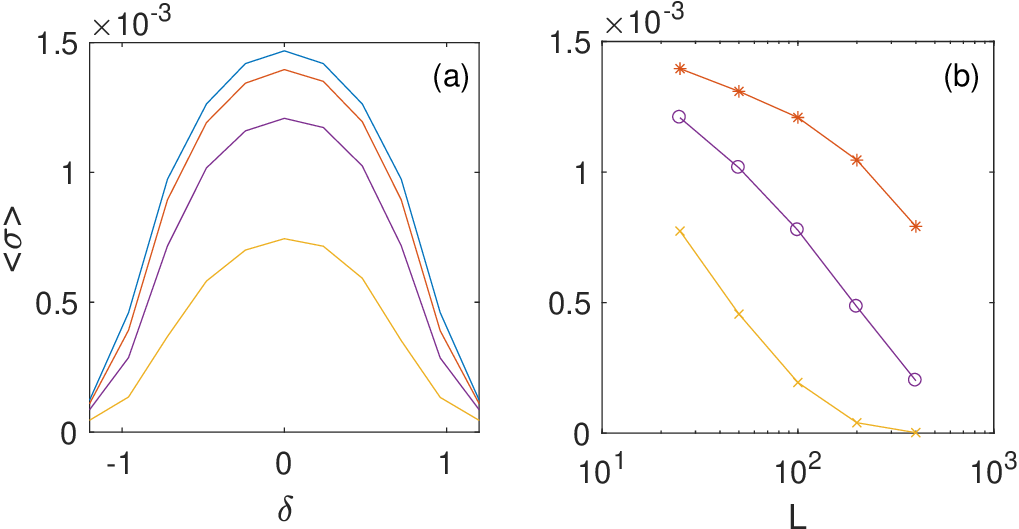}
\caption{(a) The average conductance $\bar{\sigma}=\bar{\sigma}(\delta)$ of the chain of the length $L=25$ for $\gamma=0.2$ and different disorder strength $\xi=0,0.1,0.2,0.4$. (b) The average conductance at $\delta=0$ as the function of the chain length $L=25,50,100,200,400$ for different disorder strength $\xi=0.1,0.2,0.4$.  Average over 1000 realizations of random on-site energies $\xi_\ell$.}
\label{fig6}
\end{figure}

We come to the second topic of the work. To address the Anderson localization we consider long chains up to $L=400$ and introduce the random on-site energies $\delta_\ell$,
\begin{equation} 
\delta_\ell=\delta+ \xi_\ell \;.
\end{equation} 
where $\xi_\ell$ are uniformly distributed in the interval $| \xi_\ell | \le \xi$. 
As the reference point we take the chain of the length $L=25$ whose conductance for $\xi=0$ is depicted in Fig.~\ref{fig5}(a,b). Interestingly, for disordered chains Eq.~(\ref{combined}) does not work so well as for the chains without disorder, see lower panels in Fig.~\ref{fig5}. However, on the qualitative level, it correctly captures all critical effects of the Anderson localization theory. In particular,   it  follows from the latter theory that the coefficients $\alpha_n$  are exponentially small if the chain length exceeds the Anderson localization length $L_{cr}=L_{cr}(\xi)$. Thus, the conductance also  becomes exponentially small for $L>L_{cr}$. The exact numerical analysis of the chain conductance in the presence of disorder  fully confirms this expectation, see Fig.~\ref{fig6}.

We also studied the Anderson  localization depending on the relaxation constant $\gamma$. It was found that an increase/decrease of $\gamma$ practically does not  affect the critical chain length $L_{cr}$ above which it becomes  insulating. However, an increase/decrease of $\gamma$ strongly decreases/increases fluctuations of the function $\sigma=\sigma(\delta)$. Thus, by measuring conductance fluctuations in a laboratory experiment one can extract the value of the relaxation constant $\gamma$.

\section{Dissipative lattices}

In this section we discuss the dissipative lattices which  are of particular interest for quantum transport of cold atoms \cite{Baro13,Krin17,Corm19} in optical lattice with  induced particle losses \cite{Baro13}. In general case these losses are described by the Lindblad operator
\begin{equation}
\label{loss} 
\widehat{{\cal L}}_{loss}=\sum_{\ell=1}^L\frac{\tilde{\gamma}_\ell}{2}
\left(\hat{c}_{\ell}^{\dagger}\hat{c}_{\ell}\widehat{\cal R }-2\hat{c}_{\ell}\widehat{\cal R }\hat{c}_{\ell}^{\dagger}
+\widehat{\cal R }\hat{c}_{\ell}^{\dagger}\hat{c}_{\ell} \right) \;,
\end{equation}
where   $\gamma_\ell$ is the loss rate at a given lattice cite. Playing with constants $\gamma_\ell$ one can address different physical situations, for example, to mimic ionizing  electron beam of a given width.  The limiting case where all $\gamma_\ell$  except for $\ell=\ell_0$ was analyzed in Ref.~\cite{Visu22,Visu23}. Here we  discuss the opposite situation where all constants are the same, $\gamma_\ell =\tilde{\gamma}$. 
\begin{figure}[b]
\includegraphics[width=8.5cm,clip]{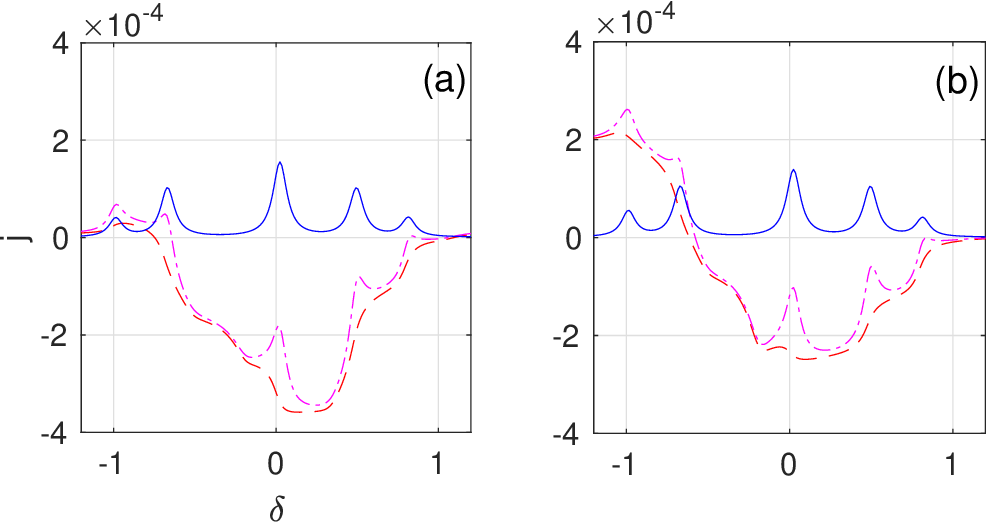}
\caption{The current Eq.~(\ref{current}) as the function of the gate voltage $\delta$ for the model (\ref{gamma_s}), left panel, and the model  (\ref{gamma_ss}), right panel. The system parameters are  $L=5$, $\xi=0.4$, $\gamma=0.1$, $\tilde{\gamma}=0.001$, and $\Delta\mu=0$ (red dashed lines) and $\Delta\mu=\pi/160$ (magenta dash-dotted lines). The blue solid lines are the net current given by the difference between dash-dotted and dashed lines.}
\label{fig}
\end{figure}

It is easy to show that adding the Lindblad operator (\ref{loss}) into the master equation (\ref{Master_full}) redefines the relaxation matrix in Eq.~(\ref{SPDM}) and Eq.~(\ref{algebraic}) as $G \rightarrow G+\widetilde{G}$, where
\begin{equation} 
\label{gamma_ss}
\widetilde{G}_{n,m}=\left\{
\begin{array}{lll}
\tilde{\gamma}&,&(n,m)\in L\times L \\
0&,&(n,m)\in M\times M \\
\tilde{\gamma}/2&,&(n,m)\in L\times M, M\times L 
\end{array}
\right. \;.
\end{equation} 
Thus, if we assume for a moment $\epsilon=0$, the chain density matrix $\rho_s(t)$ decays to zero. Commonly, the decay of off-diagonal elements of a density matrix is referred to as quantum decoherence while the decay of diagonal elements as dissipation.  It is instructive to consider these two processes separately.  We begin with the quantum decoherence without dissipation/losses where the matrix $\widetilde{G}$ has the form
\begin{equation} 
\label{gamma_s}
\widetilde{G}_{n,m}=\left\{
\begin{array}{lll}
\tilde{\gamma}(1-\delta_{n,m})&,&(n,m)\in L\times L \\
0&,&(n,m)\in M\times M \\
\tilde{\gamma}/2&,&(n,m)\in L\times M, M\times L 
\end{array}
\right. \;.
\end{equation} 
We mention, in passing, that this form of the relaxation matrix mimics to some extent decoherence effect of the phonon subsystem in solid-state devices. To distinguish this decoherence from the decoherence effect of the contacts we use the term internal decoherence.

A remarkable consequence of internal decoherence is that  it can lead to the ratchet effect \cite{Reim97,Pegu05,Hama19} where the system with broken spatial symmetry shows directed current even in the absence of chemical potential difference.  Clearly, the ratchet current is sensitive to a particular realization of the on-site disorder $\xi_\ell$ and  it is strictly zero for lattices without disorder.  As an example,  the red dashed line in Fig.~\ref{fig}(a) shows the ratchet current in the chain of the length $L=5$ for $\Delta\mu=0$ and the disorder strength $\xi=0.4$. For this particular realization of disorder the current is seen to be negative for almost all $\delta$ but for other realizations it can be positive or alternating between positive and negative values. In the case $\Delta\mu\ne0$ the ratchet current provides background for the total current, which is depicted by the dashed-dotted magenta line.  Taking the difference between the total and ratchet currents we get the net current,  see the solid blue line in Fig.~\ref{fig}(a).   We found that for $\tilde{\gamma}\ll \gamma$  the net current almost coincides with the current between the contacts in the absence of internal decoherence.  Yet, there are some deviations which become more apparent for longer lattices. Thus, the next question to address is the Anderson localization in the presence of internal decoherence.
\begin{figure}
\includegraphics[width=8.5cm,clip]{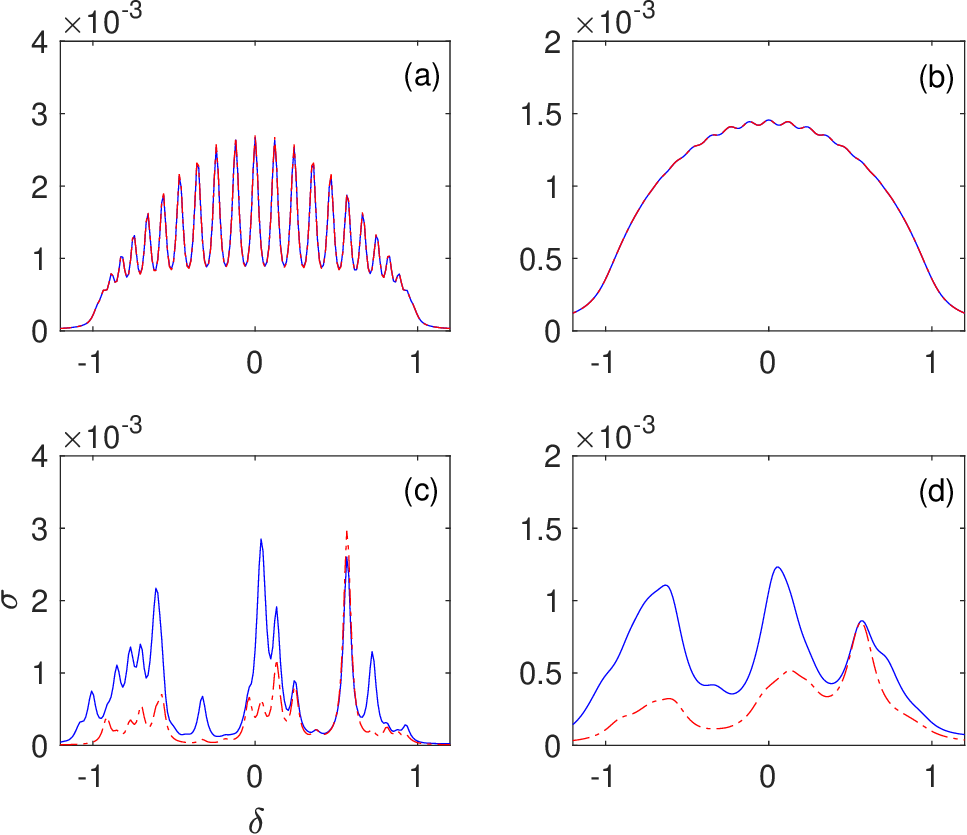}
\caption{The same as in Fig.~\ref{fig5} yet in the presence of a weak internal decoherence with the rate $\tilde{\gamma}=0.001$. The dot-dashed lines copy the solid lines from Fig.~\ref{fig5}.} 
\label{fig5e}
\end{figure}

\subsection{Anderson localization}

Since the Anderson localization is a coherent phenomenon, one may expect internal decoherence to partially destroy the Anderson localization.  We check this hypothesis by repeating the analysis of Sec.~III for  the relaxation matrix $G$ in Eq.~(\ref{algebraic}) given by the sum of the matrix (\ref{gamma}) and the matrix (\ref{gamma_s}).  As the reference point we again choose the lattice of the  length $L=25$  and restrict ourselves by the case  $\tilde{\gamma}\ll\gamma$. For $\tilde{\gamma}=0.001$ and the other parameters as in Fig.~\ref{fig5} the system conductance Eq.~(\ref{conductance}), where $j=j(\delta,\Delta\mu)$  is now the {\em net current},  is depicted in Fig.~\ref{fig5e}.  It is seen that a weak decoherence indeed enhances the conductance of the disordered chain, leaving the conductance of the perfect ($\xi=0$) chain essentially the same.

\begin{figure}
\includegraphics[width=8.5cm,clip]{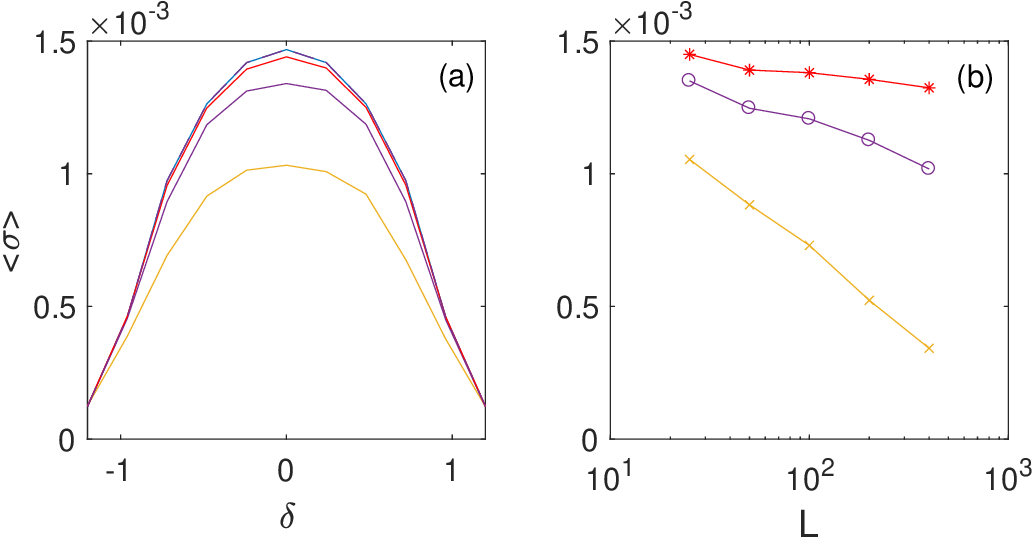}
\caption{The same as in Fig.~\ref{fig6} yet in the presence of a weak internal decoherence $\tilde{\gamma}=0.001$.}
\label{fig6b}
\end{figure}

The conductance overaged over different realizations of the on-site disorder is shown in  Fig.~\ref{fig6b}. Notice that for the average conductance we don't  need to decompose the total current into the net and ratchet currents  because the latter self-averages to zero.  Comparing the results depicted in Fig.~\ref{fig6b} with those depicted in Fig.~\ref{fig6} we conclude that a weak internal decoherence strongly suppresses the Anderson localization.  Functional  dependence of the critical chain  length $L_{cr}(\xi)$ (i.e. the length below which the disordered chain  is conducting) on the decoherence rate $\tilde{\gamma}$ remains an open problem.  

\subsection{Loss current}

Now we include dissipation into analysis, i.e., switch back to the model  Eq.~(\ref{gamma_ss}). For the chain $L=5$  the total and net currents  are shown in Fig.~\ref{fig}(b).   Having quite similar result as for the previous model, one might naively conclude that the model Eq.~(\ref{gamma_ss}) does not require a separate consideration. This is, however, not the case because, unlike for the model Eq.~(\ref{gamma_s}), the mean current Eq.~(\ref{current}) is now unrelated to the current between the contacts. Indeed, assuming for a moment  $\Delta\mu=0$, particles from both contacts go into the chain where they are lost. Thus, instead of Eq.~(\ref{current}) one should consider the loss current
\begin{equation}
\label{loss_current} 
j_{loss}=\tilde{\gamma} {\rm Tr}[\rho_s] \;,
\end{equation} 
which coincides with the sum of currents flowing from the left and right contacts.  
\begin{figure}
\includegraphics[width=8.5cm,clip]{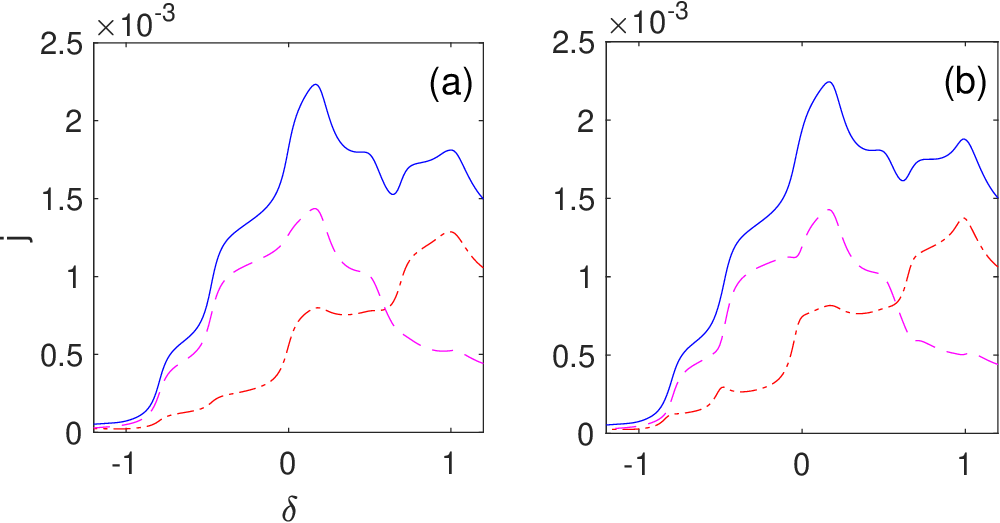}
\caption{The currents between the chain and contacts (dashed and dash-dotted lines) and the loss current (solid line) in the dissipative disordered chain. The system parameters are $L=5$, $\xi=0.4$, $\gamma=0.1$, $\tilde{\gamma} =0.001$, $\mu=0$, and $\Delta\mu=0$ (a) and $\Delta\mu=\pi/160$ (b).}
\label{fig8a}
\end{figure} 
\begin{figure}[b]
\includegraphics[width=8.5cm,clip]{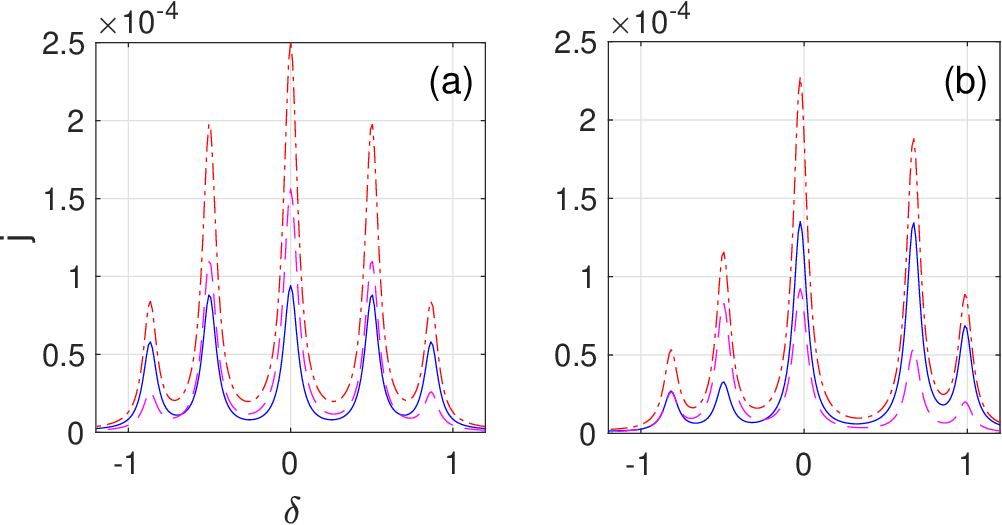}
\caption{The difference between the corresponding curves in Fig.~\ref{fig8a}. The left and right panels refer to the cases $\xi=0$ and $\xi=0.4$, respectively.}
\label{fig8b}
\end{figure} 
  
As an example, Fig.~\ref{fig8a}(a) shows these three currents for $\Delta\mu=0$ and the chain length $L=5$.  Notice that for a disordered chain the particle flow from the left and right contacts are different and they coincide only if $\xi=0$.  We also mention that the loss current is restricted from above by the value $\tilde{\gamma}L$ because for the Fermi particles ${\rm Tr}[\rho_s] \le L$. 

The right panel in Fig.~\ref{fig8a} shows the discussed three currents in the case $\Delta\mu\ne0$. At the first glance, the depicted dependences are  structureless. Yet, they carry important information which is revealed by taking the difference between the corresponding curves in the two panels, see  Fig.~\ref{fig8b}(b).  (For the sake of completeness, we also depict  in Fig.~\ref{fig8b}(a) the result for $\xi=0$.) The observed peak-like structure of the loss current is obviously a signature of the resonant transmission. Since in laboratory experiments with neutral atoms the loss current can be measured  with very high accuracy, this opens new perspectives  in studying the phenomenon of the resonant transmission.

\section{Conclusion}

We revisit the problem of two-terminal transport of non-interacting Fermi particles across the tight-binding chain by using master equation formalism.  We analyze the effect of decoherence processes in the chain (internal decoherence with the rate $\tilde{\gamma}$) and the effect of decoherence processes in contacts (external decoherence with the rate $\gamma$) on conductance of the perfect (i.e., without disorder) and disordered chains of arbitrary length $L$.  It is shown that external decoherence leads to broadening of the resonant peaks but does not affect conductance of perfect chains, which in the limit $L\rightarrow\infty$ is given by the universal equation (\ref{semicircle}).  As concerns disordered chains, external decoherence reduces conduction fluctuations but does not alter the critical length $L_{cr}$ above which the chain is insulating. 

The effect of internal decoherence is more subtle.  In the work we restricted ourselves by the case  $\tilde{\gamma}\ll \gamma$ where contribution of internal decoherence into  widths of resonant peaks and  conductance fluctuations is negligible. Yet, it is found that this weak internal decoherence drastically affects the critical length $L_{cr}$ as compared to the Anderson localization theory. 

We also consider the situation where internal decoherence is accompanied by particle losses. In this case the main object to study is the loss current,  -- number of particles which are lost in the chain per time unit. It is demonstrated that the phenomenon of resonant transmission is well reflected in the loss current.

{\em Acknowledge.} The author acknowledges fruitful discussion with D. N. Maksimov and S. V. Aksenov.


\end{document}